\documentclass[twocolumn,aps,showpacs]{revtex4}
\usepackage{amssymb}
\usepackage{amsmath,bm}
\usepackage{graphicx}
\usepackage[normalem]{ulem}
\usepackage{multirow}

\usepackage[usenames, dvipsnames]{xcolor}
\usepackage{lipsum}

\renewcommand{\sout}{\bgroup \color{red} \ULdepth=-.5ex \ULset}

\begin{document}

\title{ Light nuclei production in relativistic heavy ion collisions from the AMPT model}
\author{Kai-Jia Sun\footnote{%
kjsun@tamu.edu}}
\affiliation{Cyclotron Institute and Department of Physics and Astronomy, Texas A\&M University, College Station, Texas 77843, USA}
\author{Che Ming Ko\footnote{%
ko@comp.tamu.edu}}
\affiliation{Cyclotron Institute and Department of Physics and Astronomy, Texas A\&M University, College Station, Texas 77843, USA}

\date{\today}

\begin{abstract}
Based on an improved multiphase transport (AMPT) model, which gives a good description of proton production   with a smooth quark matter to hadronic matter transition  in relativistic heavy ion collisions, we study  deuteron  and triton production from the coalescence of nucleons at the kinetic freezeout of these collisions.  For Au+Au collisions at center-of-mass energies $\sqrt{s_\text{NN}}$ from 7.7 GeV to 200 GeV available at the Relativistic Heavy Ion Collider (RHIC), we find that the  yield ratio $N_\text{t}N_\text{p}/N_\text{d}^2$ of proton, deuteron, and triton is essentially a constant as a function of collision energy.  Our result confirms the expectation that without a first-order quark to hadronic matter phase transition in the produced matter or its approach to the critical point of the QCD matter,  this yield ratio  does not show any non-monotonic    behavior in its  collision energy dependence.
\end{abstract}

\maketitle

\section{Introduction}
\label{introduction}

Light nuclei, such as deuteron ($d$), helium-3 ($^3\text{He}$), triton ($^3\text{H}$ or $t$), helium-4 ($^4\text{He}$), hypertriton ($^3_\Lambda \text{H}$) and their antiparticles, have been observed in high energy nucleus-nucleus (AA), proton-nucleus (pA), and pp collisions at RHIC and the LHC~\cite{Abelev:2010rv,Agakishiev:2011ib,Sharma:2011ya,Adam:2019phl,Acharya:2019xmu,Acharya:2020sfy}. Because of their potential role in the search for the critical point~\cite{Bzdak:2019pkr,Luo:2020pef,Oliinychenko:2020ply} of strongly interaction matter in heavy ion collisions~\cite{Sun:2017xrx,Sun:2018jhg,Shuryak:2018lgd,Shuryak:2019ikv},
studying these loosely bound nuclei  with binding energies  much smaller than the temperature of the hot  dense matter created in these collisions has recently received an increased attention~\cite{Ma:2013xn,Andronic:2017pug,Braun-Munzinger:2018hat,Chen:2018tnh,Bazak:2018hgl,Dong:2018cye,Bellini:2018epz,Sun:2018mqq,Xu:2018jff,Cai:2019jtk,Mrowczynski:2019yrr,Kachelriess:2019taq,Mrowczynski:2020ugu,Donigus:2020ohq,Shao:2020lbq,Vovchenko:2020dmv,Ye:2020lrc,Bazak:2020wjn}. Also, these studies are useful  for understanding the production of light (anti)nuclei in  cosmic rays and in  dark matter  experiments~\cite{Vagelli:2019tqy,Kounine:2018cla,Blum:2017qnn,Poulin:2018wzu}.  As to the use of light nuclei production to probe the QCD phase diagram in relativistic heavy ion collisions, it is mainly  due to their  composite structures that make their production mostly from  nucleons close in phase space and thus sensitive to their correlations and density fluctuations 
~\cite{Sun:2017xrx,Sun:2018jhg,Shuryak:2018lgd,Shuryak:2019ikv}.
In particular, it has been suggested in Refs.~\cite{Sun:2017xrx,Sun:2018jhg} that  the yield ratio $N_\text{t}N_\text{p}/N_\text{d}^2$ of proton, deuteron, and triton in relativistic heavy ion collisions could show a non-monotonic dependence on the collision energy as a result of the enhanced density fluctuations due to the spinodal instability  during a first-order quark to hadronic matter phase transition and/or the long-range correlation  if the produced matter  is close to the critical point of the QCD matter. 

The production of light nuclei in relativistic heavy ion collisions has been studied in various  models, including the statistical hadronization model~\cite{Cleymans:1992zc,Braun-Munzinger:2015hba}, the nucleon coalescence model~\cite{Gutbrod:1988gt,Csernai:1986qf}, and dynamical models based on the kinetic theory~\cite{Danielewicz:1991dh}. In the statistical hadronization model, the yields of light nuclei  are determined at the same chemical freeze-out temperature and baryon chemical potential as those for  identified hadrons like protons, pions, kaons, etc., while their spectra are calculated  from a blast-wave model at the hadronic kinetic freeze out when  hadrons  undergo their last collisions. This model has successfully described light nuclei production in Pb+Pb collisions at $\sqrt{s_{NN}}=2.76$~TeV at LHC~\cite{Braun-Munzinger:2018hat}.  In the coalescence model, light nuclei are formed at kinetic freeze out from nucleons that are close  in phase space. There  have been various  ways of implementing the coalescence model in the literature, and these include the naive coalescence model based on the introduction of a phenomenological coalescence radius in phase space~\cite{Gutbrod:1988gt,Csernai:1986qf}, the phase-space coalescence model  that takes into account the internal wave functions of light nuclei~\cite{Sato:1981ez,Scheibl:1998tk,Mrowczynski:2016xqm,Sun:2017ooe}, and the coalescence model  that further uses the  phase-space information of nucleons from microscopic transport models~\cite{Zhu:2015voa,Dong:2018cye,Liu:2019nii,Ivanov:2017nae,Sombun:2018yqh}.  In dynamic models  for light nuclei production, these nuclei are treated as explicit degrees of freedom, and their production and destruction during the hadronic evolution stage in heavy ion collisions are described by appropriate hadronic  reactions with cross sections that satisfy the detailed balance relations~\cite{Oh:2009gx,Cho:2015exb,Oliinychenko:2018ugs}. In particular, the studies in Refs.~\cite{Cho:2015exb,Oliinychenko:2018ugs} have shown that the deuteron yield remains almost unchanged from the chemical freeze out to the kinetic freeze out as a result of the large deuteron production and destruction cross sections  that keep the deuteron abundance in thermal and chemical equilibrium in the expanding hadronic matter with decreasing temperature but increasing baryon chemical potential. All above models have been used in understanding the recent data from the STAR Collaboration on deuteron and triton in Au+Au collisions at $ \sqrt{s_{\mathrm{NN}}}$= 7.7-200 GeV~\cite{Adam:2019wnb, Zhang:2019wun,Zhang:2020ewj}.  For the thermal model, it  gives a good description of the deuteron yield  but overestimates the triton yield.    None of these models can, however,  reproduce the non-monotonic behavior or peak structure in the collision energy dependence of the yield ratio $N_\text{p}N_\text{t}/N_\text{d}^2$.

In the present work, we investigate the production of deuteron and triton in most central ($b<3~\text{fm}$) Au+Au collisions at $\sqrt{s_{\mathrm{NN}}}$ = 7.7-200 GeV by using the phase-space coalescence model based on kinetically freeze-out nucleons from an improved multiphase transport (AMPT) model~\cite{Lin:2004en,He:2017tla}. The AMPT model, which has been extensively used for studying various observables in heavy ion collisions at relativistic energies, includes the  initial conditions from the HIJING model~\cite{Gyulassy:1994ew,Wang:1991hta}, the parton cascade via the ZPC model~\cite{Zhang:1997ej}, and  the hadronic  transport based on the ART model~\cite{Li:1995pra} as well as a a spatial quark coalescence model that converts kinetically freeze-out quarks and antiquarks to the initial hadrons. Since the transition from the quark phase to the hadronic phase in AMPT is not  a first-order or second-order one, the collision energy dependence of the  yield ratio $N_\text{t}N_\text{p}/N_\text{d}^2$  from this model serves as a baseline against which experimental data can be compared to see if additional physics inputs, such as a non-smooth phase transition in the produced matter is needed in the AMPT model.  

This paper  is organized as follows. In Sec. II, we give a brief    description of  the nucleon phase-space coalescence model. We then present  in Sec. III the energy dependence of the yield ratios $N_\text{d}/N_\text{p}$, $N_\text{t}/N_\text{p}$, and $N_\text{t}N_\text{p}/N_\text{d}^2$ obtained from the coalescence model based on nucleons from the AMPT model. Finally, a conclusion is  given in Sec. V.

\begin{figure*}[t]
  \centering
  \includegraphics[width=1\textwidth]{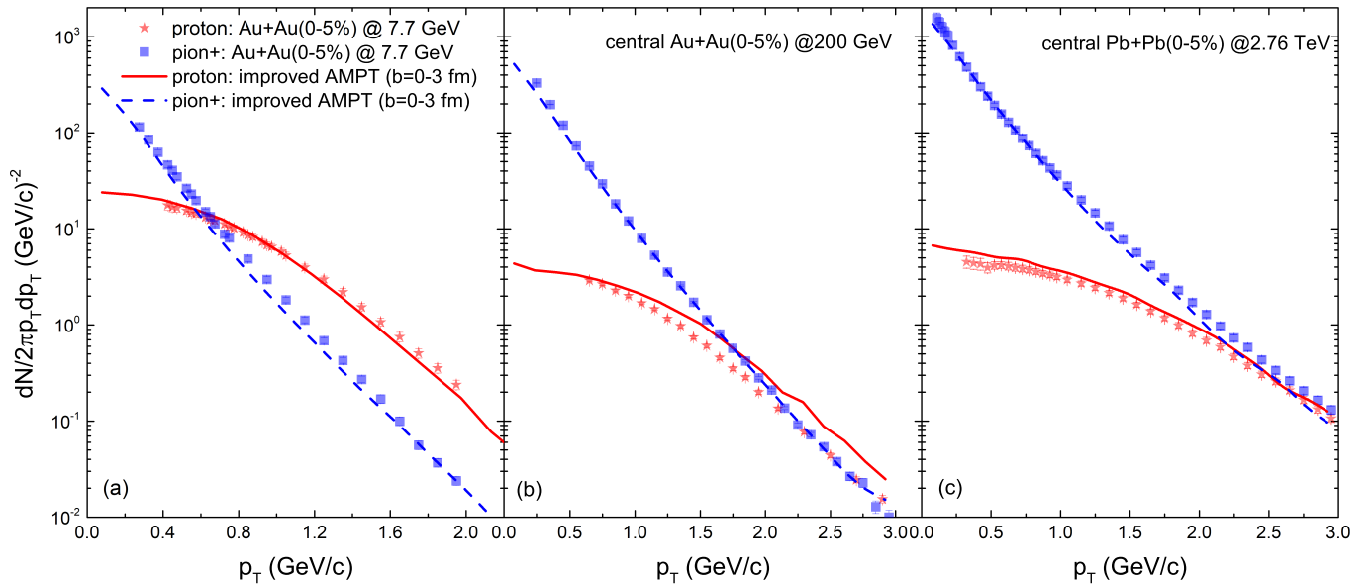}
  \caption{(Color Online)  Transverse momentum ($p_T$) spectra of protons and pions in 0-5$\%$ central Au+Au and Pb+Pb collisions at $\sqrt{s_{NN}}=$ \protect 7.7 GeV (panel (a)),  200 GeV (panel (b)),  2.76 TeV (panel (c)). Theoretical results are shown by solid lines for protons and dashed lines for pions, while the experimental  data  from Refs.~\cite{Adamczyk:2017iwn,Adler:2003cb,Abelev:2013vea} for protons and pions are denoted by  solid stars and squares, respectively.  Data for for protons at $\sqrt{s_{NN}}=$ 200 GeV and 2.76 TeV are corrected for the weak-decay contribution of hyperons.}
  \label{pic:spectra}
\end{figure*}

\section{nucleon coalescence model in phase space}

For deuteron production from an emission source of protons and neutrons, its number in the coalescence model is calculated from the overlap of the proton and neutron phase-space distribution functions $f_{p,n}({\bf x},{\bf p})$ with the Wigner function $W_{\rm d}({\bf x},{\bf p})$ of the deuteron internal wave function~\cite{Gyulassy:1982pe}, i.e., 
\begin{eqnarray}\label{d}
N_\text{d}&=&g_\text{d}\int \text{d}^3{\bf x}_1  \text{d}^3{\bf p}_1  \text{d}^3{\bf x}_2 \text{d}^3{\bf p}_2 f_n({\bf x}_1,{\bf p}_1) \notag \\
&&\times f_p({\bf x}_2,{\bf p}_2)W_\text{d}({\bf x},{\bf p}), \label{Eq3-4}
\end{eqnarray}
with $g_d=3/4$ being the statistical factor for forming a deuteron of spin 1 from two spin 1/2 proton and neutron.  Using the Gaussian or harmonic oscillator wave function for the internal wave function of a deuteron, as usually assumed in the coalescence model for deuteron production, its Wigner function is 
\begin{eqnarray}
W_\text{d}({\bf x},{\bf p})=8\exp\left({-\frac{x^2}{\sigma^2}}-\sigma^2 p^2\right)\label{Eq3-5}
\end{eqnarray}
and is normalized according to $\int \text{d}^3{\bf x}\int \text{d}^3{\bf p}~W_d({\bf x},{\bf p})=(2\pi)^3$. 
In the above, ${\bf x}$ and ${\bf p}$ are, respectively, the relative coordinate and momentum of the two nucleons in a deuteron, defined together with their center-of-mass  coordinate ${\bf X}$ and momentum ${\bf P}$ by~\cite{Chen:2003ava,Sun:2015jta,Sun:2017ooe}
\begin{eqnarray}\label{rel}
{\bf X}=\frac{{\bf x}_1+{\bf x}_2}{2},\quad{\bf x}=\frac{{\bf x}_1-{\bf x}_2}{\sqrt{2}}, \notag \\
{\bf P}={\bf p}_1+{\bf p}_2,\quad{\bf p}=\frac{{\bf p}_1-{\bf p}_2}{\sqrt{2}}.
\end{eqnarray}
The parameter $\sigma$ in Eq. (\ref{Eq3-5}) is related the root-mean-square radius $r_{\text d}$ of deuteron by $\sigma  = \sqrt{4/3}~r_{\text d}\approx 2.26$ fm and is much smaller than the size of the hadronic system considered in the present study. We note that using the more realistic Hulth{\'e}n wave function~\cite{hulthen} for the deuteron, which can be represented by 15 Guassian functions with different size parameters~\cite{Chen:2003ava}, gives essentially the same deuteron yield  because of its insensitivity to   the deuteron size in relativistic heavy-ion collisions ~\cite{Sun:2017xrx,Sun:2018jhg}. 

Similarly, the number of tritons from the coalescence of two neutrons and one proton is given by
\begin{eqnarray}\label{t}
N_\text{t}&=&g_{t}\int \text{d}^3{\bf x}_1  \text{d}^3{\bf p}_1  \text{d}^3{\bf x}_2 \text{d}^3{\bf p}_2 \text{d}^3{\bf x}_3 \text{d}^3{\bf p}_3 f_n({\bf x}_1,{\bf p}_1) \notag \\
&&\times f_n({\bf x}_2,{\bf p}_2)f_p({\bf x}_3,{\bf p}_3)W_\text{t}({\bf x},{\boldsymbol \lambda},{\bf{p}},{\bf{p}_\lambda}), \label{Eq3-6-1}
\end{eqnarray}
where $g_t=1/4$ is the statistical factor for the formation of a spin 1/2 triton from two spin 1/2 neutrons and one spin 1/2 proton. The triton Wigner function in the above equation is 
\begin{eqnarray}
W_\text{t}({\bf x},{\boldsymbol \lambda},{\bf p},{\bf p_\lambda})=8^2\exp\left({-\frac{x^2}{\sigma_\text{t}^2}-\frac{\lambda^2}{\sigma_\text{t}^2}}-\sigma_\text{t}^2 p^2-\sigma_\text{t}^2 p_\lambda^2\right),\notag \\\label{Eq3-6-2}
\end{eqnarray}
where ${\bf x}$ and ${\bf p}$ are defined as in Eq.(\ref{rel}), and $\boldsymbol\lambda$ and ${\bf p}_\lambda$ are the additional relative coordinate  and momentum.  The latter are defined together with the center-of-mass coordinate ${\bf X}$ and momentum ${\bf P}$ of the nucleons in triton by~\cite{Chen:2003ava,Sun:2015jta,Sun:2017ooe}
 \begin{eqnarray}
{\bf X}=\frac{{\bf x}_1+{\bf x}_2+{\bf x}_3}{3},~{\boldsymbol \lambda}=\frac{{\bf x}_1+{\bf x}_2-2{\bf x}_3}{\sqrt{6}}, \notag \\
{\bf P}={\bf p}_1+{\bf p}_2+{\bf p}_3, ~{\bf p_\lambda}=\frac{{\bf p}_1+{\bf p}_2-2{\bf p}_3}{\sqrt{6}}.\notag\\
\label{Eq.Jacobi}
\end{eqnarray}
The parameter $\sigma_\text{t}$ in Eq.(\ref{Eq3-6-2}) is related to the root-mean-square radius $r_\text{t}$ of triton by $\sigma_\text{t}=r_\text{t}=1.59$ fm~\cite{Ropke:2008qk}.  We note that the coordinate transformations in Eq.~(\ref{rel}) and Eq.~(\ref{Eq.Jacobi}) conserve the volume in phase space, instead of the volumes in coordinate and momentum spaces separately.

\section{Light nuclei production from the AMPT model}

In the present work, we use the improved AMPT model of Ref.~\cite{He:2017tla}, which  gives a better description of baryon production in relativistic heavy ion collisions than the usual AMPT model, to provide the phase-space information of nucleons needed for the production of deuteron and triton via the coalescence model.  Specifically, from the kinetically freeze-out nucleons given by the AMPT with each one having a freeze-out position, momentum and time, the probability for a proton and a neutron to form a deuteron is calculated from Eq.(\ref{d}) by using their relative coordinate and momentum obtained after free streaming the nucleon with an earlier freeze-out time to the later freeze-out time of the other nucleon. The similar procedure is used in calculating the probability from  Eq.(\ref{t}) for two neutrons and one proton to coalescence into a triton, i.e., by free streaming the two nucleons of earlier freeze-out times to the last freeze-out time of the remaining nucleon.

We first show in Fig.~\ref{pic:spectra}  the transverse momentum ($p_T$) spectra of protons (solid lines) and pions (dashed lines) in 0-5$\%$ central Au+Au and Pb+Pb collisions at $\sqrt{s_{NN}}=$  7.7 GeV (panel (a)),  200 GeV (panel (b)),  and 2.76 TeV (panel (c)) obtained from the theoretical calculations.  They are seen to describe fairly well the experimental data from Refs.~\cite{Adamczyk:2017iwn,Adler:2003cb,Abelev:2013vea}, shown by solid stars and squares for protons and pions, respectively, confirming the success of the improved AMPT model of Ref.~\cite{He:2017tla} in describing proton production in relativistic heavy ion collisions at both RHIC and the LHC.

\begin{figure*}[t]
  \centering
  \includegraphics[width=0.9\textwidth]{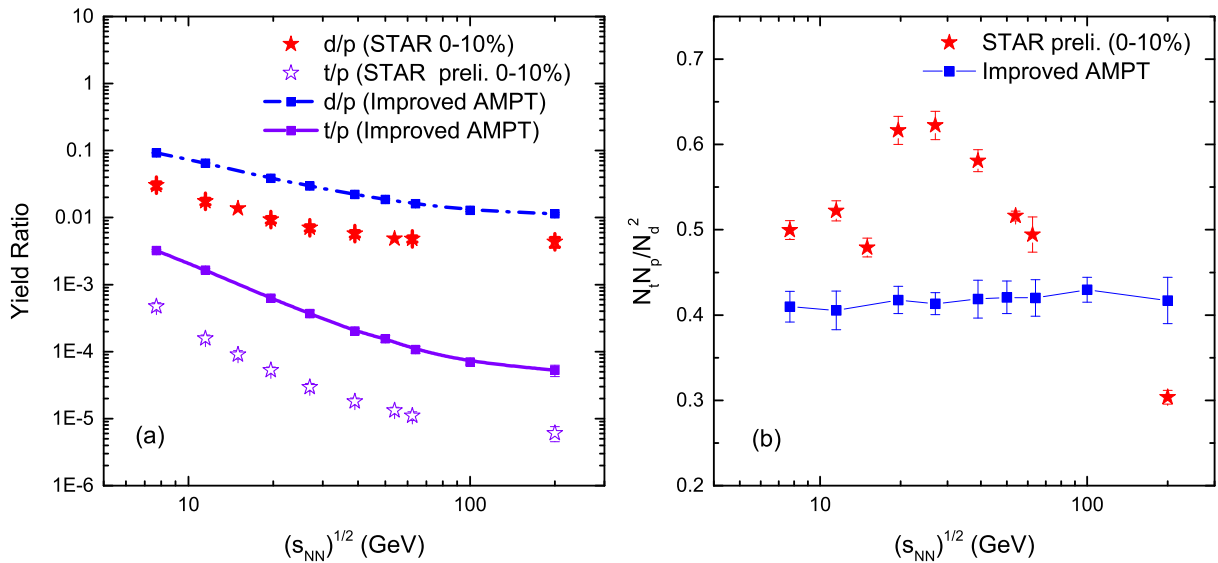}
  \caption{(Color Online) The yield ratio $N_\text{d}/N_\text{p}$  of deuteron to proton and $N_\text{t}/N_\text{p}$ of triton to proton (left window) as well as the yield ratio $N_\text{t}N_\text{p}/N_\text{d}^2$ (right window) as  functions of the collision energy $\sqrt{s_{NN}}$ in central Au+Au collisions. Results from the AMPT model are denoted by lines with solid squares, and experimental data from Refs.~\cite{Adam:2019wnb,Zhang:2019wun,Zhang:2020ewj,Liu:2019nii} are shown by solid and open stars after correcting the weak-decay contribution from hyperons to protons ~\cite{Adamczyk:2017nof}.} 
  \label{pic:ratio}
\end{figure*}

We then show in the left window of Fig.~\ref{pic:ratio}  the yield ratio $N_\text{d}/N_\text{p}$ ($d/p$) of deuteron to proton and $N_\text{t}/N_\text{p}$ ($t/p$) of triton to proton as functions of the collision energy $\sqrt{s_{NN}}$.  Results from the AMPT model, denoted by lines with solid squares, are seen  to overestimate the measured $d/p$ ratio (solid stars) by about a factor of 3 and $t/p$ ratio (open stars) by about a factor of 9 for all collision energies from 7.7 GeV to 200 GeV. The overestimation of deuteron production has also been noticed in Ref.~\cite{Zhu:2015voa} based on an earlier version of the AMPT model. This failure is related to the  low local temperature of initial hadrons at the hadronization  in AMPT, which then leads to  a similarly too low a temperature for nucleons at the kinetic freeze out. We note that for a given number of nucleons, a lower temperature results in a larger nucleon density in momentum space and thus enhances the production of deuteron and triton. Since the low local temperature is accompanied by a large radial flow in AMPT, the resulting proton transverse momentum spectrum can still reproduce the experimental data as shown in Fig.~\ref{pic:spectra}. To describe simultaneously the transverse momentum spectra of proton, deuteron and triton requires further improvements on the treatment of hadronization in AMPT, which  is, however, beyond the scope of present study.

Right window of Fig.~\ref{pic:ratio}   shows  the yield ratio $N_\text{p}N_\text{t}/N_\text{d}^2$ as a function of  the  collision energy $\sqrt{s_{NN}}$. It is seen that this ratio  has almost a constant value of around 0.4, which is slightly larger than the value of around 0.29 obtained by assuming a uniform distribution of nucleons in the analytical calculations of Refs.~\cite{Sun:2017xrx,Sun:2018jhg}.  Since this yield ratio is  insensitive to the temperature and volume of the kinetic freeze-out hypersurface, the overestimation of the  $d/p$ and $t/p$ ratios is not expected to have a large effect on this ratio.  Indeed, the overestimation factors of 3 and 9 obtained in the above for the calculated $d/p$ and $t/p$ ratios cancel exactly in the $N_\text{p}N_\text{t}/N_\text{d}^2$ ratio.  The almost collision energy independent yield ratio $N_\text{t}N_\text{p}/N_\text{d}^2$ obtained in the present study contradicts the non-monotonic behavior seen  in the data, shown by solid stars in the right window of Fig.~\ref{pic:ratio}.  This result supports the suggestion in Refs.~\cite{Sun:2017xrx,Sun:2018jhg,Shuryak:2018lgd,Shuryak:2019ikv} that a non-monotonic dependence of this ratio on the energy of heavy ion collisions requires a first-order phase transition in the produced matter or the approach to the critical point of the QCD matter.  Our result is similar to that in a recent study~\cite{Liu:2019nii}  using a pure hadronic JAM transport model with a  coalescence model based simply on some coalescence radii in phase space, where the predicted value of the yield ratio $N_\text{t}N_\text{p}/N_\text{d}^2$ is also almost a constant as a function of collision energy. The behavior of the yield ratio $N_\text{p}N_\text{t}/N_\text{d}^2$ is thus insensitive to the details of the coalescence model. Our results can therefore be used  as a baseline for understanding the background contribution in the search for the QCD critical point from heavy ion collisions in the beam energy scan (BES) program at RHIC. 

\section{conclusions}

In the present work, we  have studied light nuclei production from Au+Au collision at $\sqrt{s_\text{NN}} = 7.7~-~200$ GeV in the nucleon coalescence model  based on the phase-space distribution of nucleons from the kinetically freeze out of an improved AMPT model.  We  have found that the yield ratio $N_\text{t}N_\text{p}/N_\text{d}^2$ is essentially a constant as a function of collision energy.  Since the AMPT model only has a smooth crossover from partonic matter to hadronic matter, our results provide the baseline for understanding the background contribution to the yield ratio $N_\text{t}N_\text{p}/N_\text{d}^2$ in the BES program at RHIC.  A deviation of experimentally measured collision energy dependence of this ratio from our results could hint at the occurrence of a non-smooth phase transition in the produced matter from these collisions and thus help to determine the location of the critical point in the QCD phase diagram.

\begin{acknowledgements}
The authors thank Ziwei Lin, Xiaofeng Luo, Hui Liu, and Shanshan Cao for helpful discussions.  This work was supported in part by the U.S. Department of Energy under Contract No. DE-SC0015266 and the Welch Foundation under Grant No. A-1358.
\end{acknowledgements}

\end{document}